\documentclass[preprint,showpacs]{revtex4}
\usepackage{epsfig}
\usepackage{graphicx}
\usepackage{slashed}
\usepackage{bm}
\usepackage{amsmath} 
\usepackage{amssymb}

\newcommand{\half}{\mbox{${\textstyle \frac{1}{2}}$}}           

\newcommand{\rd}{\textrm{d}}
\begin{document}
%
\title{ Entanglement in joint $\Lambda \bar{\Lambda}$ decay; cont.}
\date{\today}
\author{G\"oran F\"aldt}\email{goran.faldt@physics.uu.se}  
\affiliation{ Department of physics and astronomy, 
Uppsala University,
 Box 516, S-751 20 Uppsala,Sweden }

\begin{abstract}
We have previously investigated joint $\Lambda \bar{\Lambda}$ decay in the reaction 
$e^+ e^- \rightarrow \gamma \Lambda(\rightarrow p\pi^-) \bar{\Lambda}(\rightarrow \bar{p}\pi^+)$.
The cross-section-distribution functions  encountered were relativistically 
covariant and expressed in terms of scalar products of the four-momentum vectors of
the  particles involved. In the present, sequel investigation, we show that by working with  
three-momentum scalars instead results could possibly become more transparent.

\end{abstract}
\maketitle
%
%
%
\section{Introduction}\label{ett}

The {\slshape BABAR} Collaboration \cite{BaBar} has  measured initial-state-radiation in
the annihilation reaction    
$e^+ e^- \rightarrow \gamma \Lambda(\rightarrow p\pi^-) \bar{\Lambda}(\rightarrow \bar{p}\pi^+)$. 
Such measurements are interesting since they offer opportunities to determine
electromagnetic form factors of the Lambda hyperons in
the time-like region. 

A theoretical analysis of this annihilation reaction is presented in 
 refs.\cite{Novo} and \cite{Czyz}. A complete determination of the cross-section-distribution 
functions, including those describing joint Lambda anti-Lambda decays, is also given in
 ref.\cite{Ent1}. The arguments of those functions, and there are several,
 are the scalar products 
of the four-momentum vectors of the particles involved. To determine all those functions 
is a formidable task, and we have therefore decided on an alternative approach,
  to work with three-momentum vectors instead. 

Replacing four-dimensional arguments by three-dimensional ones
also requires considerable work, but this work is
worth-while, as we shall see. The result is transparent. 

\section{Cross-section distribution}\label{sec2}

Our notation follows Pilkuhn \cite{Pil}. The cross-section distribution for the reaction
$e^+ e^- \rightarrow \gamma \Lambda(\rightarrow p\pi^-) \bar{\Lambda}(\rightarrow \bar{p}\pi^+)$
 is written as 
\begin{equation}
	\rd \sigma= \frac{1}{2\sqrt{\lambda(s,m_e^2,m_e^2)}} \, \overline{|{\cal{M}}|^2}\,
	   \textrm{dLips}(k_1+k_2;q,l_1,l_2,q_1,q_2)	  ,
\end{equation}	   
where the average over the squared matrix element indicates summation over final proton
and anti-proton spins and average over initial electron and positron spins.  The definitions of the
particle momenta are explained in fig.1.

We remove some trivial factors from the squared matrix element,
collected in a factor denoted ${\cal{K}}$,
\begin{equation}
	\overline{|{\cal{M}}|^2}={\cal{K}}\overline{|{\cal{M}}_{red}|^2}.\label{Msq_andK}
\end{equation}

%
\section{Previous analysis}\label{tva}

We start where our previous analysis ended, ref.\cite{Ent1}, but before we can 
do so it is necessary to
 repeat some of the important definitions  and results.

The form factors of the hyperon-electromagnetic couplings are denoted $G_1$ and $G_2$, 
a standard choice. The designations of  particle four-momenta can be  seen 
in the Feynman diagrams of fig.\ 1.
\begin{figure}[ht]
\scalebox{0.65}{\includegraphics{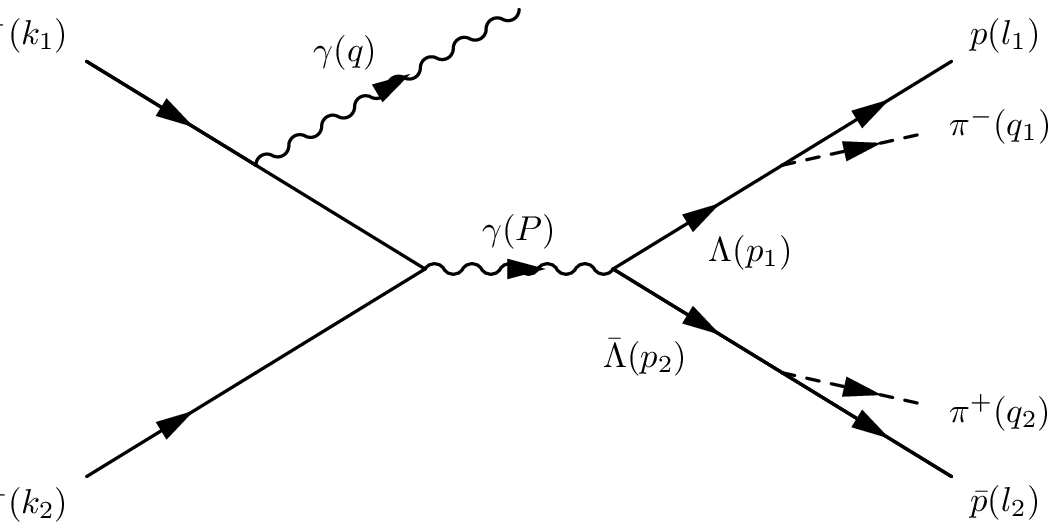}\qquad \includegraphics{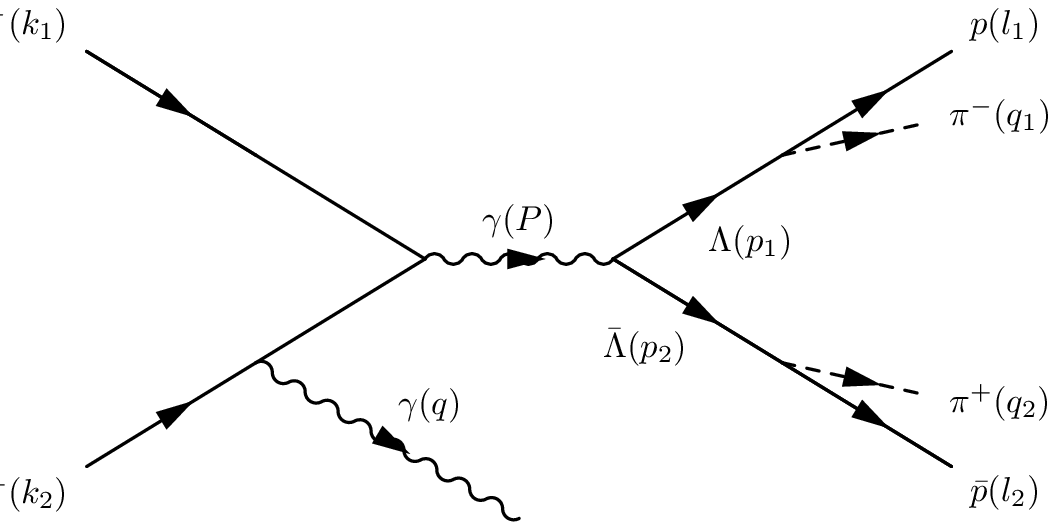}  }
\caption{Graphs included in our calculation of the reaction 
$e^+ e^- \rightarrow \gamma \Lambda(\rightarrow p\pi^-) \bar{\Lambda}(\rightarrow \bar{p}\pi^+)$.}
\label{F1-fig}
\end{figure}

The cross-section-distribution function, or rather the covariant square of the annihilation 
matrix element $\overline{|{\cal{M}}_{red}|^2}$, 
 is the contraction of hadronic $H_{\mu\nu}$
 and leptonic $L^{\mu\nu}$ tensors, 
\begin{equation}
	\overline{|{\cal{M}}_{red}|^2}=L^{\mu\nu}H_{\mu\nu}.
\end{equation}
 Its right-hand side is naturally decomposed as
\begin{equation}
	\overline{|{\cal{M}}_{red}|^2} =\bar{R}_\Lambda R_\Lambda M^{RR}+\bar{R}_\Lambda S_\Lambda M^{RS}
	   +\bar{S}_\Lambda R_\Lambda M^{SR} +\bar{S}_\Lambda S_\Lambda M ^{SS},\label{MM-decomp}
\end{equation}
with suffixes $RS$ and $\bar{R}\bar{S}$ referring  to $\Lambda$ and $\bar{\Lambda}$ decay constants, 
with $R$ the spin-independent  and $S$ the spin-dependent one. 

From the structure of the lepton tensor,  eq.(24) of ref.\cite{Ent1}, one concludes that 
each of the $M^{XY}$ functions has two parts,
\begin{equation}
	M^{XY}=-a_y A^{XY}(G_1,G_2) -b_yB^{XY}  (G_1,G_2) ,\label{MXY}
\end{equation}
where the $A^{XY}$ factor is obtained by contracting the hadron tensor 
with the symmetric tensor
$	k_{1\mu}k_{1\nu} + k_{2\mu}k_{2\nu}$, 
and the $B^{XY}$ factor by contraction with the tensor $ g_{\mu\nu}.$ 
For details see ref.\cite{Ent1}. The weight factors 
$a_y$ and $b_y$ are defined in appendix A.

The functions $A^{XY}$ and $B^{XY}$ are bilinear forms of $G_1$ and $G_2$, and we expand
them accordingly, for $A^{XY}$,
\begin{eqnarray}
A^{XY}(G_1,G_2) &=&|G_1|^2 {\cal{K}}^{AXY}_1 + |G_2|^2 {\cal{K}}^{AXY}_2\nonumber\\
	  &+&2 \Re (G_1G_2^{\star}) {\cal{K}}^{AXY}_3 +2 \Im (G_1G_2^{\star}) {\cal{K}}^{AXY}_4,   
				\label{ABdef}
\end{eqnarray}
and similarly for $B^{XY}$.
We refer to  the functions $\{ {\cal{K}}\} $ as co-factors. 
They are Lorentz covariant functions of the particle four-momenta
and the functions of our attention.
 
%
%
\section{Previous results}\label{sectfem}

The leading term of eq.(\ref{MM-decomp}) is  $M^{RR}$ as it is independent of 
variables that relate to spin dependences in 
the hyperon decay distributions. We have
\begin{eqnarray}
	A^{RR} &=& 2|G_1|^2\bigg[ (k_1\cdot P)^2+(k_2\cdot P)^2-(k_1\cdot Q)^2-(k_2\cdot Q)^2\bigg] \nonumber\\
	&& +4\Re(G_1G_2^\star) \bigg[(k_1\cdot Q)^2+(k_2\cdot Q)^2\bigg]  
	 - |G_2|^2 \frac{Q^2}{2M^2}\bigg[(k_1\cdot Q)^2+(k_2\cdot Q)^2\bigg] , \label{eqARR}
\end{eqnarray}
with $Q=p_1-p_2.$ Furthermore,  
\begin{eqnarray}
	B^{RR} &=& -4|G_1|^2( P^2+2M^2)
	 +4\Re(G_1G_2^\star) Q^2  -|G_2|^2 \frac{(Q^2)^2}{2M^2}.\label{eqBRR}
\end{eqnarray}
Thus, the distribution function $M ^{RR}$ does not depend on the decay momenta 
$l$ and $q$ of the Lambda hyperons.

Next in order are terms linear in the spin variables,
\begin{eqnarray}
	A^{RS} &=& -4\Im(G_1G_2^\star)  
	   \bigg[   k_1\cdot Q\, \mbox{det}(p_2p_1l_1k_1) +  k_2\cdot Q\, \mbox{det}(p_2p_1l_1k_2)\bigg]  ,\label{EARS}\\
	  A^{SR} &=& -4 \Im(G_1G_2^\star)  
	   \bigg[   k_1\cdot Q\, \mbox{det}(p_2p_1l_2k_1) +  k_2\cdot Q\, \mbox{det}(p_2p_1l_2k_2)\bigg] ,\label{EASR}
\end{eqnarray}
with $\mbox{det}(abcd)=\epsilon_{\alpha\beta\gamma\delta}a^\alpha b^\beta c^\gamma d^\delta$ and 
\begin{eqnarray}
	B^{RS}& =& 0, \label{eqBRS} \\	B^{SR} &= &0\label{eqBSR}	.
\end{eqnarray}

The expressions for the spin-spin contributions are more complicated.
 We have for the $A$ contribution
\begin{eqnarray}
	A^{SS} &=& - 2|G_1|^2 \bigg[   \bigg((k_1\cdot P)^2+(k_2\cdot P)^2-(k_1\cdot Q)^2-(k_2\cdot Q)^2\bigg)
	 \bigg(p_1\cdot l_1 p_2\cdot l_2+M^2l_1\cdot l_2\bigg) \nonumber \\
	 && \qquad \quad +2M^2\bigg(  P^2(k_1\cdot l_1 k_1\cdot l_2  +k_2\cdot l_1 k_2\cdot l_2) 
	  -2P\cdot l_2(k_1\cdot l_1 k_1\cdot p_2 +k_2\cdot l_1 k_2\cdot p_2 )\nonumber \\
	&& \qquad \qquad\qquad\quad -2P\cdot l_1(k_1\cdot l_2 k_1\cdot p_1 +k_2\cdot l_2 k_2\cdot p_1 )\bigg)
	  \bigg] \nonumber \\
	  &&-4\Re(G_1G_2^\star)\bigg[ M^2l_1\cdot l_2\bigg( (k_1\cdot Q)^2+(k_2\cdot Q)^2 \bigg)
	  \nonumber \\
	 && \qquad \quad - M^2\bigg( P\cdot l_2(k_1\cdot Q k_1\cdot l_1 + k_2\cdot Q k_2\cdot l_1 ) 
	 -P\cdot l_1(k_1\cdot Q k_1\cdot l_2 + k_2\cdot Q k_2\cdot l_2 )\bigg)  \nonumber \\
	 &&\qquad \quad -k_1\cdot Q\bigg(k_1\cdot p_1 p_2\cdot l_1p_2\cdot l_2-k_1\cdot p_2 p_1\cdot l_1p_1\cdot l_2-
	   \half P^2 k_1\cdot l_1p_2\cdot l_2 +\half P^2 k_1\cdot l_2 p_1\cdot l_1  \bigg)
		\nonumber \\
	 &&\qquad \quad -k_2\cdot Q\bigg(k_2\cdot p_1 p_2\cdot l_1p_2\cdot l_2-k_2\cdot p_2 p_1\cdot l_1p_1\cdot l_2-
	   \half P^2 k_2\cdot l_1p_2\cdot l_2 +\half P^2 k_2\cdot l_2 p_1\cdot l_1  \bigg) \bigg]\nonumber \\
	   &&-|G_2|^2 \frac{1}{2M^2}\bigg( (k_1\cdot Q)^2+(k_2\cdot Q)^2\bigg)\bigg[ Q^2
	     \bigg( p_1\cdot l_1p_2\cdot l_2-M^2l_1\cdot l_2\bigg)
	     +2M^2Q\cdot l_1 Q\cdot l_2 \bigg]  , \label{eqASS}
\end{eqnarray}
and for the $B$ contribution 
\begin{eqnarray}
	B^{SS} &=&+4|G_1|^2 \bigg[  (P^2+2M^2)(p_1\cdot l_1 p_2\cdot l_2 +M^2 l_1\cdot l_2) \nonumber \\
	 && \qquad\qquad  -M^2\bigg( P^2l_1\cdot l_2 +2 P\cdot l_2 l_1\cdot p_1
	       +2 P\cdot l_1 l_2\cdot p_2\bigg)\bigg]     \nonumber \\
	  &&  -4\Re(G_1G_2^\star)\bigg[ Q^2 M^2 l_1\cdot l_2 -
	     M^2\bigg( Q\cdot l_1P\cdot l_2-Q\cdot l_2P\cdot l_1 \bigg)     \nonumber \\
	 &&   \qquad \quad -\bigg( p_1\cdot Q p_2\cdot l_1 p_2\cdot l_2 -   p_2\cdot Q   p_1\cdot l_1  p_1\cdot l_2
	        -\half P^2 Q\cdot l_1 p_2\cdot l_2 +\half P^2 Q\cdot l_2 p_1\cdot l_1             \bigg) \bigg] \nonumber \\
	&&  -|G_2|^2 \frac{Q^2}{2M^2} \bigg[Q^2
	     \bigg( p_1\cdot l_1p_2\cdot l_2-M^2l_1\cdot l_2 \bigg)
	     +2M^2Q\cdot l_1 Q\cdot l_2 \bigg].  \label{eqBSS}
\end{eqnarray}

The functions $A^{SS}$ and $B^{SS}$ describe the joint-decay distributions of the Lambda and anti-Lambda 
hyperons. The distributions are
correlated, {\itshape i.e.},
 they cannot be written as a product of Lambda and 
anti-Lambda distribution functions. 
Our distribution functions are explicitly covariant, as they are
expressed in terms of the four-momentum vectors of the participating particles. It is not necessary  to 
work in several coordinate systems, as in refs.\ \cite{BaBar} and \cite{Czyz}. Another important point is that our
calculation correctly counts  the number of intermediate hyperon states.
%
%
%
  \section{Reference frames}\label{reffr}	
The cross-section distribution function of
sect.	\ref{sectfem} involves expressions that are  
functions of scalar products of  particle four-momenta. To determine the scalar product of two
	four-vectors requires knowledge of those vectors in one and the same 
	reference frame. Our task in this section is to demonstrate
	how this is achieved.
	
	Designations of the particle four-momenta follow from 
	the energy-momentum-balance condition in
	the reaction $e^+e^-\rightarrow \bar{\Lambda}(\rightarrow\bar{p}\pi^+)
	 \Lambda(\rightarrow p\pi\-)\gamma$, 
	\begin{equation}
	 k_1+ k_2=p_1+p_2+q.
	\end{equation}
	Additional information is contained  in fig.\ 1.

	The gamma three-momentum $\mathbf{q}$, and electron
	three-momentum $\mathbf{k}$, are  momenta defined
	in the $e^+e^-$ centre-of-momentum (c.m.) reference frame, 
	in which  $\hat{\mathbf{q}} \cdot \hat{\mathbf{k}} =\cos \theta$.
	We refer to this  frame as $S_0$.
	 In $S_0$ electron
	and positron four-momenta are 
		$k_1 = \epsilon(1, \hat{\mathbf{k}})$ and
		$k_2 = \epsilon(1, - \hat{\mathbf{k}}) $,
	with $\epsilon $ the common lepton energy. 
With  $\omega$ the gamma  energy,  the gamma four-momentum is denoted 
 $q=\omega(1, \hat{\mathbf{q}})$. Furthermore, the four-momenta
of  Lambda
and anti-Lambda 
are  $p_1=(E_1, \mathbf{p}_1)$ and $p_2=(E_2, \mathbf{p}_2)$.

Now, we shall not perform our calculations  in $S_0$ but in $S_1 $
which is the c.m.\ frame of the $\Lambda\bar{\Lambda}$ pair. 
We indicate variables in this frame by a prime, so that 
\begin{equation}
p_{1,2}^\prime = (E_{\Lambda},\pm p_{\Lambda}\mathbf{f}) \label{Lamf}
\end{equation}
with $E_{\Lambda}=\sqrt{p_\Lambda^2+M_{\Lambda}^2}$ and $\mathbf{f}$ a unit vector. 
The $\Lambda\bar{\Lambda}$ c.m.\  energy $W=2E_{\Lambda}$ may be obtained from the 
identity
\begin{equation}
W^2=4\epsilon(\epsilon-\omega).
\end{equation}

The next question concerns the relation between  frames
 $S_1$ and $S_0$. Since 
$\mathbf{p}_1+\mathbf{p}_2=-\mathbf{q}$ in $S_0$, we argue that  $S_1$  can
be reached from $S_0$ through a boost 
along the direction of motion of the gamma, and of 
magnitude,
\begin{equation}
	v=
	\frac{-(\mathbf{p}_{1} + \mathbf{p}_{2} )\cdot\mathbf{q}}{E_1+E_2}
	=\frac{\omega}{\sqrt{\omega^2+W^2}}, 
	\label{velo}
\end{equation} 
and with Lorentz-transformation (LT) coefficient
	\begin{equation}
	\gamma(v)=\frac{1}{\sqrt{1-v^2}} 
	=\frac{\sqrt{\omega^2+W^2}}{W} .
\end{equation}
 Also, note that $v$
is the relative velocity between two reference frames, it is not a particle velocity.

A Lorentz boost from  $S_0$ to $S_1$ leads to 
new four-momentum vectors for the initial state leptons, namely 
\begin{equation}
k_{1,2}^\prime=\epsilon \gamma \bigg[ (1\pm v \mathbf{n}\cdot \hat{\mathbf{k}});\
 v(\mathbf{n} \pm \mathbf{N})\bigg], \label{S1K}
\end{equation}
and with $\mathbf{n}$ and $\mathbf{N}$  by definition
\begin{eqnarray}
\mathbf{n}&=& \hat{\mathbf{q}},\label{S3K}\\
\mathbf{N}&=&\frac{1}{ v\gamma} \bigg[\hat{\mathbf{k}}+(\gamma -1)
(\mathbf{n}\cdot\hat{\mathbf{k}})\mathbf{n}\bigg].\label{DefN}
\end{eqnarray}
Relations (\ref{S3K}) and (\ref{DefN}) are identical to 
those introduced by the BaBar collaboration \cite{BaBar}.

The photon radiated in our annihilation  process  carries energy $\omega$ and   three-momentum $\mathbf{q}=\omega \mathbf{n}$ ,
 when observed in $S_0$. A boost  from $S_0$ to $S_1$, sends this vector into
$\mathbf{q}^\prime = \omega^\prime \mathbf{n}$,  with
\begin{equation}
\omega^\prime = \omega \sqrt{\frac{1+v}{1-v}}.  
\end{equation}

However, we should not  forget the decay products of the hyperons, 
the antiproton and the proton.
 In the rest system $S_2$ of the Lambda
the proton is
represented by the four-vector 
\begin{equation}
l_1=(E_g, p_g \, \mathbf{g}),
\end{equation}
with $\mathbf{g}$  a unit vector, and with decay parameters 
$p_g$ and $E_g$.

 Similarly, in the rest system $S_3$ of the anti-Lambda 
 the anti-proton is represented
by the four-vector
\begin{equation}
l_2=(E_h, p_h \mathbf{h}),
\end{equation} 
with $\mathbf{h}$  a unit vector, and 
decay parameters $p_h=p_g$ and $E_h=E_g$. 
A passage from $S_3$
to $S_1$ is achieved by a Lorentz boost with velocity  $v_h$ 
and direction $\mathbf{f}$, whereas 
a passage from $S_2$
to $S_1$ is achieved by a Lorentz boost with velocity  $v_g$ 
and direction $-\mathbf{f}$.

The boost equations for the massive hyperons are well known. Vectors 
orthogonal to the boost velocity, $\mathbf{v}=v\mathbf{n}$, are unchanged, 
those parallel are changed according to the Lorentz-transformation 
prescription 
 \begin{eqnarray}
	\mathbf{p}_{1,2}^\prime &=& \big[ \mathbf{p}_{1,2} -
	\mathbf{n} (\mathbf{n}\cdot \mathbf{p}_{1,2})\big] +\gamma(v)
	\mathbf{n} \big[ 
\mathbf{n}\cdot\mathbf{p}_{1,2} +v E_{1,2}\big], \label{LTek}\\
		E_{1,2}^\prime &=& \gamma(v)\big[ E_{1,2} -
		\mathbf{v}\cdot\mathbf{p}_{1,2} \big].\label{LTe}
\end{eqnarray}
The inverse-transformation equations, going from  $S_1$ to 
$S_0$, are obtained by changing the sign of the velocity, 
from $\mathbf{n}$  to 
$-\mathbf{n}$.

After this elementary discussion we are ready for the proton and antiproton 
four-momentum vectors in $S_1$; for the proton
\begin{eqnarray}
	\mathbf{p}_g^\prime &=& 
	\gamma_\Lambda E_g \mathbf{f}  \big[ 
	  \mathbf{f}\cdot ( \mathbf{v}_{g} + \mathbf{v}_\Lambda ) \big]+
		\mathbf{p}_{g\bot} ,
			\label{gtrans}\\
	E_g^{\prime }&=& \gamma_\Lambda E_g\big[1 + \mathbf{v}_\Lambda\cdot \mathbf{v}_{g}
\big]  ,
\end{eqnarray} 
 the transverse-vector component $\mathbf{p}_{g\bot}$ being  
\begin{equation}
\mathbf{p}_{g\bot}=\mathbf{p}_{g}-
		(\mathbf{p}_{g}\cdot \mathbf{f})\, \mathbf{f},
\end{equation}
with $\mathbf{p}_{g\bot}\cdot\mathbf{f} =0;$ 
 and for the antiproton
\begin{eqnarray}
	\mathbf{p}_h^\prime &=& 
	\gamma_\Lambda E_h \mathbf{f}  \big[ 
	  \mathbf{f}\cdot ( \mathbf{v}_{h} - \mathbf{v}_\Lambda ) \big]+
		\mathbf{p}_{h\bot} ,
		\label{htrans}	\\
	E_h^{\prime }&=& \gamma_\Lambda E_h\big[1 - \mathbf{v}_\Lambda\cdot \mathbf{v}_{h}
\big]  .
\end{eqnarray} 
Our calculations make use of the shorthand notations, 
\begin{align}
	 G_g&= (\mathbf{v}_{g}+\mathbf{v}_\Lambda) \cdot \mathbf{f}, &  
	H_g&=1+ \mathbf{v}_\Lambda \cdot \mathbf{v}_g ,\\ 
	G_h&= ( \mathbf{v}_{h} -\mathbf{v}_\Lambda)\cdot \mathbf{f}, &
	H_h&=1- \mathbf{v}_\Lambda \cdot \mathbf{v}_h  .
\end{align} 
%
%
%
	%
	%

%
%

 
  \section{Calculating co-factors} \label{co-fac}
  Co-factors can be identified in the $A^{XY}$ and
the $B^{XY}$  functional distributions of sect.\ \ref{sectfem}. 
The results are 
co-factors expressed in terms of scalar products of four-vector momenta. 
However, our goal was to find simpler expressons, and this by
evaluating all scalars in one and the same reference frame, 
the c.m. reference frame  of the $\Lambda \bar{\Lambda}$ pair.

	We have evaluated cross-section distributions for
	two sets of form-factor parameters. The 
	two sets have attached form-factor sets, that we
	indicate by different letters, such that 
	$(G_1,G_2)\Rightarrow \{ {\cal{K}} \}$ 
	and $(G_M,G_E)\Rightarrow \{ {\cal{L}} \}$ .
	We start with the ${\cal{K}}$ set and return to the 
	${\cal{L}}$ set in sect.\ref{KapGeGm}.
	
	The much needed $\Omega$ functions are defined in appendix B.
	 The spin-correlation functions are equally important. Their
	definitions are, 
	\begin{eqnarray}
		X_a(\mathbf{g},\mathbf{h} )& =&  2 
		\mathbf{g} \cdot\mathbf{f}\, \mathbf{h} \cdot\mathbf{f} 
		- \mathbf{g}\cdot\mathbf{h}   ,\label{Xa} \\
			X_b(\mathbf{g},\mathbf{h} )& =& 
			\mathbf{g} \cdot\mathbf{f}\, \mathbf{h} \cdot\mathbf{f}.\label{Xb}
	\end{eqnarray}

	A parameter that appears in practically every formula is the $Z$
	 parameter,
	\begin{equation}
		Z=\frac{4M_{\Lambda}^2}{Q^2}=	
		\frac{-1}{\gamma_{\Lambda}^2 v_{\Lambda}^2}=1-\frac{1}{v_{\Lambda}^2}.
		\label{defZETA}
	\end{equation}
Other important parameters are $v_\Lambda =p_\Lambda / E_\Lambda$ 
and 
$\gamma_\Lambda=E_\Lambda /M_\Lambda$.      
%

	%
	%
\subsection{Spin-independent co-factors}

We start with the ${\cal{K}}$ base and the ${ \cal{K}}^{ARR}$ co-factors,
which can be extracted from
the $A^{RR}$ contribution to the   $M^{RR}$ functional distribution of
 eq.(\ref{eqARR}). Since he calculation is straightforward
we are satisfied with the result
\begin{eqnarray}
		{\cal{K}}_1^{ARR} &=&( 2 \epsilon \omega)^2\Omega_f v_{\Lambda}^2
		 \bigg[ -Z+\Omega_{\bot}/(\Omega_f  v_{\Lambda}^2) \bigg], \\
				{\cal{K}}_2^{ARR}& =&(2\epsilon \omega)^2 \Omega_f v_{\Lambda}^2
				 \bigg[\frac{-1}{ Z} \bigg],\\
						{\cal{K}}_3^{ARR} &=&(2\epsilon \omega)^2\Omega_f v_{\Lambda}^2 
		\bigg[ 1\bigg].	
	\end{eqnarray}
	The $\Omega$ functions are described in appendix B.

Next, we extract the ${\cal{K}}^{BRR}$ co-factors 
from  the $B^{RR}$ contribution (\ref{eqBRR}) to the   $M^{RR}$ functional distribution  of eq.(\ref{MXY});
\begin{eqnarray}
		{\cal{K}}_1^{BRR}& =& - 2Q^2\bigg[ Z-2(1-Z) \bigg] , \\
				{\cal{K}}_2^{BRR}& =& - 2Q^2\bigg[ \frac{1}{Z}\bigg], \\
						{\cal{K}}_3^{BRR}& =&- 2Q^2\bigg[ -1\bigg].
	\end{eqnarray}
	
	Knowledge of these co-factors leads immediately  to the   
	$A^{RR}$ and $B^{RR}$ functions of eq.(\ref{ABdef}),
	\begin{eqnarray}
		A^{RR} &=&( 2 \epsilon \omega)^2  \frac{1}{Z(1-Z)} 
		 \bigg[ -|Z G_1 -G_2|^2\Omega_f +Z(1-Z) 
		   |G_1|^2 \Omega_{\bot} \bigg], \\
		 B^{RR} &=& \frac{2Q^2}{Z}
		 \bigg[ |Z G_1 -G_2|^2+ 2Z(1-Z) |G_1|^2
				 \bigg],
	\end{eqnarray}
	expressions which are well-known and also displayed
	in refs.\cite{Novo}
	and   \cite{Czyz}.   
	
	
\subsection{Linearly spin-dependent co-factors}	

Next in order are terms linear in the spin variables, represented by the
 functions of eqs.(\ref{EARS})
and (\ref{EASR}), and their only non-vanishing co-factors ${\cal K}^{ARS}_4$ 
and    ${\cal K}^{ASR}_4$. 
The expessions for the prefactors of the above-mentioned  equations 
are quite easily obtained, and equals  
\begin{equation}
Q\cdot k_{1,2}= -\frac{2p_\Lambda\epsilon\omega}{W}
\mathbf{f} \cdot( \mathbf{n}\pm\mathbf{N}).
\end{equation}
In the $S_1$ frame the determinant boils down to
	\begin{equation}
		\mbox{det}(p_2^\prime p_1^\prime l_1^\prime k_2^\prime)= 2E_\Lambda \mathbf{p}_1^\prime \cdot 
		 (\mathbf{p}_{g}^\prime \times \mathbf{k}_2^\prime )
		=2E_\Lambda p_\Lambda
		\mathbf{f} \cdot 
		 (\mathbf{p}_{g}^\prime\times \mathbf{k}_2^\prime ).
	\end{equation}
	Hence, components of the vectors  $\mathbf{p}_{g}^\prime$
	or $\mathbf{k}_2^\prime $ along $\mathbf{f}$ will not 
	contribute to the value of the  determinant, so that by eq.(\ref{gtrans}) we may replace 
	$\mathbf{p}_{g}^\prime$ by $\mathbf{p}_{g}$.
 For $\mathbf{k}_2^\prime$ we make recourse to eq.(\ref{S1K}). 
All this yields,
\begin{eqnarray}
	{\cal K}^{ARS}_4 &=& 2p_{\Lambda }p_g (2\epsilon  \omega)^2 
	v_{\Lambda}
	\bigg[ \mathbf{f}\cdot \mathbf{n} 
	\mathbf{f} \cdot ( \mathbf{g} \times \mathbf{n} ) +
	\mathbf{f}\cdot \mathbf{N}\mathbf{f} \cdot ( \mathbf{g} \times \mathbf{N} )
	\bigg]  ,\label{KRS}\\
	  {\cal K}^{ASR}_4 &=&  2p_{\Lambda }p_g (2\epsilon  \omega)^2 
	v_{\Lambda}
	\bigg[ \mathbf{f}\cdot \mathbf{n} 
	\mathbf{f} \cdot ( \mathbf{h} \times \mathbf{n} ) +
	\mathbf{f}\cdot \mathbf{N}\mathbf{f} \cdot ( \mathbf{h} \times \mathbf{N} )
	\bigg] .
	  \label{KSR}
\end{eqnarray}
The co-factors that relate to $B^{XY}$ of eq.(\ref{MXY}) vanish,
\begin{eqnarray}
	{\cal K}^{BRS}_4 &=&0,\label{BRSa}\\
	  {\cal K}^{BSR}_4 &=&  0.
	  \label{BSRa}
\end{eqnarray}

With the co-factors of eqs.(\ref{KRS}) and (\ref{KSR})    in hand we can determine $A^{RS}$ and $A^{SR}$ from 
eq.(\ref{ABdef}).  The related  functions $B^{RS}$ and $B^{SR}$ vanish identically. In agreement with  refs.\cite{Novo},     
\begin{eqnarray}
 A^{RS} &=& 2p_{\Lambda }v_\Lambda p_g (2\epsilon  \omega)^2 
	2\Im(G_1G_2^{\star})
	\bigg[ \mathbf{f}\cdot \mathbf{n} 
	\mathbf{f} \cdot ( \mathbf{g} \times \mathbf{n} ) +
	\mathbf{f}\cdot \mathbf{N}\mathbf{f} \cdot ( \mathbf{g} \times \mathbf{N} )
	\bigg]  ,\label{ARS}\\
 A^{SR}&=& 2p_{\Lambda }v_\Lambda p_g (2\epsilon  \omega)^2 
	2\Im(G_1G_2^{\star})
	\bigg[ \mathbf{f}\cdot \mathbf{n} 
	\mathbf{f} \cdot ( \mathbf{h} \times \mathbf{n} ) +
	\mathbf{f}\cdot \mathbf{N}\mathbf{f} \cdot ( \mathbf{h} \times \mathbf{N} )
	\bigg] .
	  \label{ASR}
\end{eqnarray}
%
%
\subsection{Doubly spin-dependent co-factors}\label{st5.c}

Now, the co-factors are suffixed $ASS$ and  $BSS$. Those 
 suffixed $ASS$ are obtained by analysing   $A^{SS}$ of eq.(\ref{eqASS});
\begin{eqnarray}
		{\cal{K}}_1^{ASS} &=& (2 p_\Lambda  p_g \epsilon \omega)^2 
		 \Omega_f v_{\Lambda}^2\,  
		\bigg[  -Z^2 X_a
  + 		\frac{Z\Omega_{\bot}}{v_{\Lambda}^2 \Omega_f} \big(X_a-2X_b )+
	\frac{Z}{v_{\Lambda}^2 \Omega_f}
	 B_1 \bigg], \\	
				{\cal{K}}_2^{ASS}& =& 
									(2 p_\Lambda  p_g \epsilon \omega)^2 
		 \Omega_f  v_{\Lambda}^2
					\bigg[ - X_a \bigg] , \\
					{\cal{K}}_3^{ASS} &=&(2 p_\Lambda  p_g \epsilon \omega)^2 
		 \Omega_f  v_{\Lambda}^2
					\bigg[ ZX_a  +\frac{Z}{v_{\Lambda}^2 \Omega_f}B_3\bigg] , 
	\end{eqnarray}
	with  functions $B_1$ and $B_3$ as defined in 
	appendix \ref{appB}. The co-factors  suffixed $BSS$ are dug out  
	from  $B^{SS}$ of eq.(\ref{eqBSS});
\begin{eqnarray}
		{\cal{K}}_1^{BSS} &=& 2(2 p_\Lambda^2 p_g)^2    \,  
		\bigg[ 
		Z^2 X_a+ 2Z(1-Z) X_b \bigg] , \\	
	{\cal{K}}_2^{BSS}& =& 2(2 p_\Lambda^2 p_g)^2  \,\bigg[ 
				 X_a\bigg], \\
			{\cal{K}}_3^{BSS} &=& 2(2 p_\Lambda^2 p_g)^2  \, 
					\bigg[-Z X_a\bigg],
	\end{eqnarray}
	where $Z$ is defined in eq.(\ref{defZETA}). 

	Starting from the ${\cal{K}}^{ASS}$ and ${\cal{K}}^{BSS}$ 
	co-factors we easily derive
	functions $A^{SS}$ and $B^{SS}$, 
	\begin{eqnarray}
		A^{SS} &=& 4(p_\Lambda p_g  \epsilon \omega)^2  
		 \bigg[ -|Z G_1 -G_2|^2 \frac{\Omega_f}{1-Z} X_a+
		 Z\Omega_\bot |G_1|^2(X_a-2X_b)	
		\nonumber \\
		&&+2Z\Re (G_1G_2^{\star})B_3
		 +Z|G_1|^2B_1 \bigg] ,
		\\
		 B^{SS} &=& 8 (p_\Lambda^2 p_g)^2 
		 \bigg[ |Z G_1-G_2|^2 X_a +2Z(1-Z) |G_1|^2 X_b
				 \bigg],	
	\end{eqnarray}
	two functions which have not been investigated before. Now,  
	we have all the ingrediants needed to calculate the 
	cross-section-distribution function frm 
	eqs.(\ref{MM-decomp}-\ref{ABdef}).
  \section{The $G_E/G_M$ set}\label{KapGeGm}

%

	Up to  now, we have only considered an expansion of  the cross-section-distribution functions $A^{XY}$ and $B^{XY}$ in terms of 
 the form factors $G_1$ and $G_2$  and  
their co-factors. Other choices of form factors are possible and we 
shall in particular consider the pair $G_E$ and $G_M$. The two sets
are related by 
\begin{eqnarray}
	G_1&=& G_M	 , \label{FFM}\\
	G_2&= &\frac{4M_\Lambda^2}{Q^2}(G_M -G_E). \label{FFE}
\end{eqnarray}
 The arguments of the form factors are all equal to $P^2$. In particular, when 
$P^2=4M_\Lambda^2$ then $G_M=G_E$.

The functions $A^{XY}$ and $B^{XY}$ are bilinear forms of $G_1$ and $G_2$, and expanded
according to eq.(\ref{ABdef}), but they can also be expanded in terms 
of $G_M$ and $G_E$ in which case 
\begin{eqnarray}
A^{XY}(G_M,G_E) &=&|G_M|^2 {\cal{L}}^{AXY}_1 + |G_E|^2 {\cal{L}}^{AXY}_2\\
	  &+& 2 \Re (G_MG_E^{\star}) {\cal{L}}^{AXY}_3 + 2\Im (G_MG_E^{\star}) {\cal{L}}^{AXY}_4,   
				\label{CDdef}
\end{eqnarray}
and similarly for $B^{XY}$. 
The relation between the two sets of functions, \{${\cal{K}}_i^{XYZ}$\} 
	and \{${\cal{L}}_i^{XYZ}$\}  becomes
	\begin{eqnarray}
		{\cal{L}}_1 &=& { \cal{K}}_1  + Z^2{\cal{K}}_2 + 2Z
	       {\cal{K}}_3, \\
	{\cal{L}}_2 &=&  Z^2{\cal{K}}_2,  \\
	{\cal{L}}_3 &=&- Z^2{\cal{K}}_2 - Z{\cal{K}}_3\\
	{\cal{L}}_4 &=&- Z{\cal{K}}_4,
	\end{eqnarray}
	and with the parameter $Z$
	defined in eq.(\ref{defZETA}). The most notible fact 
	of the new set is that several  co-factors vanish;
	\begin{equation}
	{\cal{L}}_3^{ARR}={\cal{L}}_3^{BRR}={\cal{L}}_3^{BSS}=0,
	\end{equation}
	whereas ${\cal{L}}_3^{ASS}\neq 0.$
	
	Co-factors suffixed ${ARR}$;
	\begin{eqnarray}
		{\cal{L}}_1^{ARR} &=&(2 \epsilon \omega)^2\,  \bigg[ (\mathbf{n}  \times\mathbf{f})^2
		+ ( \mathbf{N} \times\mathbf{f})^2 \bigg], \\
				{\cal{L}}_2^{ARR}& =&(2 \epsilon \omega)^2 \, \frac{1}{ \gamma_{\Lambda}^2}\,
				\bigg[ ( \mathbf{n}  \cdot\mathbf{f})^2
		+ ( \mathbf{N} \cdot\mathbf{f})^2 \bigg] ,  \\
						{\cal{L}}_3^{ARR} &=&0.
	\end{eqnarray}
	Co-factors suffixed ${BRR}$; 
\begin{eqnarray}
		{\cal{L}}_1^{BRR}& =& -4 P ^2   , \\
				{\cal{L}}_2^{BRR}& =& 8M^2, \\
						{\cal{L}}_3^{BRR}& =&0.
	\end{eqnarray}
Co-factors suffixed ${ARS}$  and ${ASR}$;
\begin{eqnarray}
	{\cal L}^{ARS}_4 &=& -2Zp_{\Lambda }p_g (2\epsilon  \omega)^2 
	v_{\Lambda}
	\bigg[ \mathbf{f}\cdot \mathbf{n} 
	\mathbf{f} \cdot ( \mathbf{g} \times \mathbf{n} ) +
	\mathbf{f}\cdot \mathbf{N}\mathbf{f} \cdot ( \mathbf{g} \times \mathbf{N} )
	\bigg]  ,\label{KRS2}\\
	  {\cal L}^{ASR}_4 &=& - 2Zp_{\Lambda }p_g (2\epsilon  \omega)^2 
	v_{\Lambda}
	\bigg[ \mathbf{f}\cdot \mathbf{n} 
	\mathbf{f} \cdot ( \mathbf{h} \times \mathbf{n} ) +
	\mathbf{f}\cdot \mathbf{N}\mathbf{f} \cdot ( \mathbf{h} \times \mathbf{N} )
	\bigg] .
	  \label{KSR2}
\end{eqnarray}
The co-factors that relate to $B^{XY}$ of eq.(\ref{MXY}) vanish,
\begin{eqnarray}
	{\cal K}^{BRS}_4 &=&0,\label{BRS}\\
	  {\cal K}^{BSRb}_4 &=&  0.
	  \label{BSRb}
\end{eqnarray}

With these co-factors in hand we can determine $A^{RS}$ and $A^{SR}$ from 
Eq.(\ref{ABdef}) whereas the related  functions $B^{RS}$ and $B^{SR}$ vanish.
%

%
%
%
  \section{Discussion}
  
	In a previous report, ref.\cite{Ent1}, we derived a set of  cross-section-distribution 
	functions for the reaction  
$e^+ e^- \rightarrow \gamma \Lambda(\rightarrow p\pi^-) \bar{\Lambda}(\rightarrow \bar{p}\pi^+)$. 
It involved functions whose arguments were Lorentzian scalar products
of four-momentum vectors of the particles participating in the reaction.
Since some of the functons are quite intricate we have here tried another 
approach, replacing the Lorentz scalars by Euclidean scalars of three-momentum
vectors. In doing so it should be  remembered that one can form a scalar product 
of two vectors only if they are defined in the same reference system. 
The functions which multiply the coupling constants  are called co-factors, or weight factors,  
 and can be retrieved  from eqs.(\ref{eqARR}-\ref{eqBSS}). For a 
complete determination of the cross-section-distribution 
function fourteen co-factors are needed.

The cross-section-distribution function describing our  $e^+e^-$ 
annihilation reaction is detailed in sect.\ \ref{co-fac}. However, 
we are also interested in  
knowing the final-state-distribution function 
after integrating over one or 
both hyperon decays. Let us start with the integral over the $\Omega_g$
hyperon decay, while keeping the anti-hyperon decay angle $\Omega_h$ fixed. 
Then, terms which are linear in the three-vector $\mathbf{g}$  vanish. Thus,
\begin{equation}
\int \frac{\rd \Omega_g}{4\pi} A^{SS},B^{SS},A^{RS}=0.
\end{equation}
Also, since $B^{RS}$ is non-excisting we can add $B^{RS}=0$ leaving the three-vector 
$\mathbf{h}$ described by the co-vector of  eq.(\ref{KSR}). 

The next step is integration over the hyperon decay angles $\Omega_h$,
\begin{equation}
\int \frac{\rd \Omega_h}{4\pi} A^{SR}=0.
\end{equation}
Since  by definition, $B^{SR}=0$, we end up with the cross-section-distribution 
function for the reaction
$e^+ e^- \rightarrow \gamma \Lambda\bar{\Lambda}$, as expected. 
This distribution function is proportional to the function 
$M^{RR}$ of eq.(\ref{MM-decomp}).

The angular integrations just described can also be performed in the 
four-dimensional formulation of the co-factors,  as 
 in sect.\ \ref{sectfem}, 
by exploiting eq.(7.48) of  
ref.\cite{Ent1} for the integration.

There is an alternative approach to  the angular integration \cite{Eul},  which 
employs the  Euler angles. In this case the angular measure is 
written as
\begin{equation}
\rd \Omega_h \rd \Omega_g   = \rd (\cos\theta_{gh}\big) 
\rd \alpha \rd (\cos\beta) \rd \gamma . 
\end{equation}
Since we know that terms linear in the vectors  $\mathbf{g}$ or $\mathbf{h}$
vanish upon angular integration, we need only concern ourselves 
with the co-factors of $A^{SS}$ and $B^{SS}$ of sect.\ \ref{st5.c}.
We notice that $\cos\theta_{gh}$ only appears in the function
$X_a(\mathbf{g},\mathbf{h} )$ of eq.(\ref{Xa}), 
\begin{equation}
X_a(\mathbf{g},\mathbf{h} )  = 2 
		\mathbf{g} \cdot\mathbf{f}\, \mathbf{h} \cdot\mathbf{f} 
		- \mathbf{g}\cdot\mathbf{h} .\nonumber
\end{equation}		
The first term in this expression vanishes on the 
$\alpha,\gamma$ integrations, 
and $\mathbf{g} \cdot\mathbf{h}=\cos\theta_{gh}$.

With a little help from the co-factors of sect.\ \cite{Eul} 
we get closed expressions for $B^{SS}$ and $A^{SS}$, 
  \begin{align}
  B^{SS} &= 8\cos\theta_{gh} (p_\Lambda^2 p_g)^2  
  \Big(  -|Z G_1-G_2|^2  \Big) ,\\
  A^{SS}&= 4 \cos\theta_{gh} (p_\Lambda p_g\epsilon
	\omega)^2 \frac{1}{1-Z}\Big(
  |Z G_1-G_2|^2  \Omega_f  +Z(1-Z)|G_1|^2\Omega_{\bot} \Big).  
 \end{align}
Both functions vanish on integration
 over the $\cos\theta_{gh}$ variable. Moreover, $ZG_1-G_2=ZG_E$
   and  $G_1=G_M$.

%
%
\appendix
\section{Kinematics explained}\label{AppA}

The parameters describing the decay of Lambda  into  proton and  pion are $p_g$ and $E_g$, with 
\begin{align}
	p_g&=\frac{1}{2M_\Lambda} \bigg[ \big( (M_\Lambda +m_p)^2-\mu^2\big) 
	   \big( (M_\Lambda-m_p)^2-\mu^2\big) \bigg]^{1/2} ,\\
	E_g  &=\frac{1}{2M_\Lambda}\big(M_\Lambda^2 +m_p^2-\mu^2\big) ,
\end{align}
representing the proton in the Lambda rest system.

The kinematic variables $P^2,$ $y_1, $ and $y_2$ of 
eq.(\ref{MXY}) are defined by,
\begin{eqnarray}
P^2 &=&(p_1+p_2)^2 ,\\
	 y_1 &=&2k_1    \cdot q=2\epsilon\omega(1-\cos\theta),\\
	  y_2 &=&2 k_2\cdot q =2\epsilon\omega(1+\cos\theta),        
\end{eqnarray}
and the normalisation factors $a_y$ and $b_y$ of the same equation by
\begin{eqnarray}
a_y &=&4P^2/(y_1y_2) ,\\
	 b_y &=&(2sP^2+y_1^2+y_2^2)/(y_1y_2).
\end{eqnarray}
%
\section{Notations explained}\label{appB}

The  $\Omega$	fuctions are by definition, and $\mathbf{n}^2=1$,
\begin{eqnarray}
		\Omega(\mathbf{n},\mathbf{N})&=&\mathbf{n}^2+ \mathbf{N}^2
		=\frac{1}{v^2}+ (  \mathbf{n} \cdot \hat{\mathbf{k}})^2,\\
		&=&\Omega_f(\mathbf{n},\mathbf{N}) +
		\Omega_{\bot}(\mathbf{n},\mathbf{N}), \\
		\Omega_f(\mathbf{n},\mathbf{N})&=&
		 (\mathbf{n}  \cdot\mathbf{f})^2
		+ ( \mathbf{N} \cdot\mathbf{f})^2, \\
				\Omega_{\bot}(\mathbf{n},\mathbf{N})&=& (\mathbf{n}  \times\mathbf{f})^2
		+ ( \mathbf{N} \times\mathbf{f})^2.
	\end{eqnarray}

 The  $\mathbf{N}$ vector is  defined in eq.(\ref{DefN}), and 
fulfils  the relations,
	\begin{eqnarray}
	\mathbf{N}^2&=&\frac{W^2}{\omega^2} +(\mathbf{n}\cdot \hat{\mathbf{k}})^2,\\
	\mathbf{n}\cdot\mathbf{N}&=& \frac{2\epsilon-\omega}{\omega}
   \mathbf{n}\cdot \hat{\mathbf{k}}.
	\end{eqnarray}

	Also, introduced are co-factor functions $B_1$ and $B_3$
\begin{eqnarray}
		L_0 &=&\mathbf{n}\cdot\mathbf{g}_\bot \,  \mathbf{n}\cdot \mathbf{h}_\bot +
		\mathbf{N} \cdot
		\mathbf{g}_\bot \, \mathbf{N} \cdot \mathbf{h}_\bot ,\\
		L_M &=& (\mathbf{f}\cdot\mathbf{g} \mathbf{h}_\bot +
			\mathbf{f}\cdot\mathbf{h} \mathbf{g}_\bot )\cdot
			( \mathbf{n}\mathbf{f}\cdot\mathbf{n}  +
			\mathbf{N}\mathbf{f}\cdot\mathbf{N})
			,\\
			B_1 &=& 2\bigg[ L_0 +\frac{1}{\gamma_\Lambda}
			L_M \bigg] ,\\
			B_3&=&\gamma_\Lambda L_M,
	\end{eqnarray} 
	with $\gamma_\Lambda=E_\Lambda/M_\Lambda$.      
%

%

\end{document}